\documentclass[aps,prl,twocolumn,showpacs]{revtex4}
\usepackage{amsmath}
\usepackage{epsfig}
\usepackage{amssymb}
\usepackage{dcolumn}
\usepackage{graphicx}
\usepackage{dcolumn}

\usepackage{color}

\def\dd#1#2{{{\rm d} #1\over {\rm d} #2}}

\def\bb#1{{\bf #1}}

\newcommand{\beq}{\begin{equation}}
\newcommand{\eeq}{\end{equation}}

\begin{document}
\date{\today}

\title{Transitions and excitations in a superfluid stream passing small impurities}

\author{ Florian Pinsker${}^1$ and Natalia G. Berloff${}^{1,2}$}
\email{florian.pinsker@gmail.com}

\affiliation{${}^1$Department of Applied Mathematics and 
Theoretical Physics, University of Cambridge, Wilberforce Road, CB3 0WA, United Kingdom.\\
${}^{1,2}$Cambridge-Skoltech Quantum Fluids Laboratory,  Skolkovo Institute of Science and Technology
ul. Novaya, d.100,
Skolkovo 143025 Russian Federation}

\begin{abstract}
We analyse asymptotically and numerically the motion around a single impurity  and  a network of impurities inserted in a two-dimensional superfluid. The criticality for the break down of superfluidity is shown to occur when it becomes energetically favourable to create a doublet -- the limiting case between a vortex pair and a rarefaction pulse on the surface of the impurity. Depending on the characteristics of the potential representing the impurity different excitation scenarios are shown to exist for a single impurity as well as for a lattice of impurities. Depending on the lattice characteristics it is shown that several regimes are possible: dissipationless flow, excitations emitted by the lattice boundary, excitations created in the bulk and the formation of  large scale structures.
\end{abstract}


\maketitle

\section{Introduction}

Superfluidity is the property of extraordinary low viscosity in a fluid for which evidence was first found in Liquid Helium II, i.e., He-4 below the $\lambda$-point at $2.17$K \cite{kap,mdi}. Later it was proposed that Liquid Helium II  can be regarded as a degenerated Bose-Einstein gas in the lowest energy mode \cite{London} and the hydrodynamical picture was completed by arguing that the condensed fraction \cite{E} of the superfluid does not take part in the dissipation of momentum \cite{Lon}; it is due to the non-condensed atoms/molecules or quasi particles that viscosity occurs in superfluids. In 1995 condensation to the lowest energy state was achieved experimentally for weakly interacting dilute Bose-gases \cite{first,second,third}  and subsequently  research on hydrodynamic properties has been presented; superfluidity could be confirmed by moving laser beams of different shapes through the condensate, which showed a drag force only above some critical velocity and dissipationless flow below \cite{evid, ketter,heat, barrier}. Experimental investigations have been accompanied by a great advancement in our theoretical understanding of this matter; a variety of scenarios \cite{Gross}, like superfluid flow around obstacles of different shapes at sonic or supersonic speed \cite{frisch, pom,ham, win, spies, pita, glad,berloff,Pavloff, wat}, solitary waves due to inhomogeneities \cite{dark, comp, train} or the transitions emerging in rotating Bose gases \cite{Fe1,Fe2,Fe3,Pin,Pin2} have been considered. In more recent years superfluids made out of quasiparticles such as gaseous coupled Fermions (Cooper pairs) \cite{pair,fete}, Spinor condensates \cite{Spinor,spinor2,spinor3}, Exciton-Polaritons \cite{supi2,supi,exit,polariton,berloff3} or classical waves \cite{classic,classic2,Fleisch} have received much attention \cite{Light,leg}. New aspects emerge for investigation and the quest of elucidating defining properties of these novel superfluids goes on \cite{berloff4}.  

To study the nature of a superfluid, in particular the key property of zero viscosity at zero temperature $T=0$, a well-established scenario to consider is an obstacle in relative motion to the fluid \cite{evid, ketter, heat, frisch, pom,ham, win, spies, pita, glad,berloff,Pavloff, wat}. It was noted as early as in 1768 by d'Alembert \cite{grim} that an incompressible and inviscid potential flow past an obstacle encounters vanishing resistance, i.e., such a fluid is in a state of superfluidity. In this paper, however, we shall consider a {\it compressible} superfluid with zero viscosity obeying a finite speed of sound, which is to be described by a nonlinear Schr\"odinger-type equation. Among such superfluids governed by nonlinear Schr\"odinger-type equations are dilute and weakly interacting Bose-gases \cite{Lieb}, condensates of classical waves \cite{Fleisch} or Exciton-Polariton condensates close to equilibrium \cite{berloff3}. Due to the finite speed of sound the fluid obeys a critical velocity $v_{\text{cr}}$ above which excitations occur within the condensate; superfluidity starts to dissipate, nodal points in the condensate wave function (with zero absolute value and discontinuities in the phase) might emerge forming quantized vortices \cite{Bruch}. 
{ In the presence of an obstacle this criterion needs to be modified as the full depletion of the condensate on the surface means that the local speed of sound is zero, this however does not lead to the excitation formation at small velocities.  It has been shown by numerical simulations  that this transition takes place when it becomes energetically favourable for a vortex to appear inside the healing layer on the surface of the obstacle \cite{huepe00}.} By generating such elementary excitations as vortices or rarefaction pulses { (for smaller perturbations)} the system limits the superfluid flow velocity by the onset of a drag force on the moving obstacle due to those excitations. 
Emerging rarefaction pulses in the wake of the obstacle may exchange energy due to sound waves (or more generally perturbations) propagating through the condensate or via contact interaction as they encounter other excitations. When acquiring energy rarefaction pulses  may transform into vortex pairs (or vortex rings in settings of $3$ spatial dimensions) or by radiating energy vortex pairs collapse to excitations of lower energy \cite{berloff6}. 

In this paper we consider a condensate in $2$ spatial dimensions ($2$d) passing finite size obstacles that are about the size of the superfluid's healing length. At the very first the obstacles (which  are assumed to be repulsive) either induce vortex pairs or rarefaction pulses into the superfluid's stream; weaker impurities imposing just a slight dip of small radius on the superfluid density support the generation of rarefaction pulses while stronger interacting impurities favor the appearance of a vortex pair. Experimentally it has been shown in \cite{BAnderson} that vortex pairs are stable excitations in oblate and effectively $2$d Bose-Einstein condensates.

Using a species-selective dipole potential the  localized impurities were created in experiment \cite{catani}. To elucidate what kind of flows can exist in such systems we consider many impurities arranged on specific lattices and demonstrate that various regimes are possible depending on the network configuration. The motion of generated excitations within the lattice is strongly affected by their attraction to the impurities and interactions between excitations. 

Our paper is organized as follows.  In Section 1 we use asymptotic expansions to estimate the density,  velocity and energy  of the condensate moving past a fixed obstacle and determine analytically the speed at which the vortex nucleation takes place as a function of the radius of the obstacle.  In Section 2 we analyse how the form of the potential modelling the impurity affects the excitations nucleated. In Section 3 we study various regimes in superfluid flow passing impurity networks. We conclude by summarizing our findings.

 \section{1. Asymptotic expansion for a  flow around a disk below the criticality}

In this section we extend the analysis done in \cite{berloff} for the flow around a two-dimensional disk of a radius large compared to the healing length to obtain the corrections due to the finite disk size. 
We study the superfluid flow in $2$d; the condensate order parameter satisfies the Gross-Pitaevskii equation  \cite{reviews}  in the reference frame moving with velocity $v$,
\begin{equation}\label{modelone}
i \hbar \partial_t \psi = - \frac{\hbar^2}{2 M} \Delta \psi + V^{\rm nat} \psi - (E + \frac{1}{2} M v^2 - g |\psi|^2) \psi.
\end{equation} 
The expression $E + \frac{1}{2} M v^2$ is the energy in the moving reference frame at which the impurity is at rest. In this model the superfluid consists of $N$ Bosons each with mass $M$, $g>0$ is the strength of the repulsive self-interactions within the fluid and dependent on the number of its atoms \cite{Inge}, $V^{\rm nat}$ denotes the potential modeling the impurities, $E$ the single-particle energy in the laboratory frame. A natural scale in our discussion is the healing length, $\xi = \hbar/(2 M E)^{1/2}$.

We develop the asymptotics of the state function $\psi = \sqrt{\rho} \exp(i \phi)$. We use the nonlinear Schr\"odinger equation \eqref{modelone} where, for simplicity, we drop the potential $V^{\rm nat} $ in favor of boundary conditions on the wave function for which the potential is an impenetrable barrier of radius $b$, so $\psi(r<b,t)=0$, where $r^2=x^2+y^2$. We write Eq.  \eqref{modelone} in hydrodynamical form using the Madelung transformation

\begin{equation}
\psi = R{\rm e}^{iS},
\label{mt}
\end{equation}
so that
\begin{equation}
\rho = M R^2, 
\qquad \phi = (\hbar/M)S,
\label{rjp}
\end{equation}
and rescale the resulting equations using $\vec x \to b \vec x$, $t \to (\xi b M/\hbar) t$, $v \to (\hbar/\xi M) U$ and $\psi \to \psi_\infty \psi$ and using the dimensionless parameter $\varepsilon = \xi/b$. The resulting system of equations  becomes 
\begin{eqnarray}
&\varepsilon^2\nabla^2R - R(\nabla S)^2 = (R^2 - 1- U^2)R
\label{aamain1}\\
&R\nabla^2S + 2\nabla R \cdot \nabla S = 0,
\label{aamain2}
\end{eqnarray}
subject to the boundary conditions $R=0$ at $r=1$, $S\rightarrow -U x$ and $R\rightarrow 1$ as $r\rightarrow \infty$. We shall assume that both $\varepsilon$ and $U$ are small and  consider an asymptotic expansion   of the solution to Eqs. \eqref{aamain1} and \eqref{aamain2}. 

{\it Boundary layer,-} At the boundary layer the quantum pressure contribution plays a crucial role. We introduce $r=1 + \varepsilon \chi$ and  expand $R$ and $S$ as
\beq\label{ser}
R(\chi,\theta) = \widehat R_0 (\chi,\theta) + \varepsilon \widehat R_1 (\chi,\theta) + \varepsilon^2 \widehat R_2(\chi,\theta) + \ldots
\eeq
\beq\label{ser2}
S(\chi,\theta)= \widehat S_0 (\chi,\theta) + \varepsilon \widehat S_1(\chi,\theta)+ \varepsilon^2 \widehat S_2(\chi,\theta) + \ldots
\eeq
The  solutions  to $O(\varepsilon^2)$ in $\hat S$ and the leading order for $R$ were found in \cite{berloff} for a spherical object, for the disk these become
\beq
\widehat R_0 = g(\theta) \tanh\left(g(\theta) \chi/ \sqrt{2} \right), \quad \widehat S_0 = \widehat S_0 (\theta), \quad \widehat S_1 = \widehat S_1 (\theta)
\eeq

\begin{equation}
\widehat{S}_2 = -{\partial \over \partial \theta}
\biggl(h(\chi,\theta) \dd {\widehat{S}_0} \theta\biggr) + 
\zeta_2(\theta),
\label{S2}
\end{equation}
where
\begin{eqnarray}
h(\chi,\theta) &=& \int_0^\chi {d\chi' \over \widehat{R}_0^2(\chi',\theta)}
\int_0^{\chi'} \widehat{R}_0^2(\chi{''},\theta)d\chi'' \nonumber\\
&=& {1 \over 2} 
\chi^2 - {\sqrt{2}\chi\over g(\theta)}{\rm coth}
\Bigl({g(\theta)\chi \over \sqrt{2}}\Bigr),
\label{h}
\end{eqnarray}
and $\hat S_0 (\theta), \hat S_1 (\theta)$ and $ \zeta_2(\theta)$ are functions  that will be 
determined by matching to the mainstream and  
\beq
g(\theta) = \sqrt{\left(1+ U^2 - \left(\hat S'_0 (\theta) \right)^2 \right)}.
\eeq

{\it Mainstream,-}  To leading order, the mainstream flow is governed by 
\beq\label{ma}
R^2 = 1+U^2 - (\nabla S)^2
\eeq
\beq
R^2 \nabla^2 S + \nabla R^2 \nabla S =0
\eeq
that can be combined to a single equation on $S$
\begin{equation}
(1+U^2 - 3(\nabla S)^2)\nabla^2 S =0.
\label{criterion}
\end{equation}
We expand $S$ in powers of $\varepsilon$ as in  \cite{berloff}
\begin{equation}
S(\vec x, t)=S_0(\vec x, t) + \varepsilon S_1(\vec x, t)+ \varepsilon S_2(\vec x, t)+ \cdot\cdot\cdot
\end{equation}
where $S_0, S_1$ etc. are expanded in powers of $U$ as 
\begin{equation}
S_0 = U S_{11}(r) \cos\theta + U^3(S_{31}(r)\cos\theta 
+ S_{33}(r)\cos3\theta) + \cdot\cdot\cdot, 
\label{S2d}
\end{equation}
where we assumed that  $\theta = 0$ is parallel to $\bb v$. The solutions for the mainstream we find up to $O(U^{11})$; the first few  are 

\begin{eqnarray}
S_{11}&=&-\frac{r^2+1}{r}, \quad S_{31}=\frac{c_1}{r}+\frac{6 r^2-1}{6 r^5}, \quad S_{33}=\frac{c_2}{r^3}+\frac{1}{2 r},\nonumber \\
S_{51}&=&\frac{c_1}{2 r^5}-\frac{2 {c_1}}{r^3}+\frac{{c_2}}{2 r^7}-\frac{{c_2}}{r^5}+\frac{{c_3}}{r}-\frac{7}{30 r^9} ,\nonumber \\
&+&\frac{19}{12
   r^7}-\frac{8}{3 r^5}+\frac{3}{2 r^3}\nonumber\\
   S_{53}&=&-\frac{{c_1}}{2 r}+\frac{3 {c_2}}{5 r^7}-\frac{3 {c_2}}{r^5}+\frac{{c_4}}{r^3}-\frac{1}{36 r^9}+\frac{11}{30
   r^7}, \nonumber \\
   &-&\frac{2}{r^5}+\frac{1}{2 r}\nonumber\\
   S_{55}&=&-\frac{3 {c_2}}{2 r^3}+\frac{{c_5}}{r^5}-\frac{3}{4 r^3}-\frac{1}{4 r}.
\end{eqnarray}
To carry out the asymptotic matching, we substitute $r=1 + \varepsilon\xi$ into the mainstream functions, expand the solution \eqref{S2d} in powers of $\varepsilon$ and match it to the boundary layer solution. To this order it is the same as to request that $S_{ij}'(r)=0$ at $r=1$, so on the boundary of the disk. Thus we found the solution $S_0$ to $O(U^{11})$. The first few terms are (correcting the expression given in \cite{berloff})
\begin{widetext}
\begin{align*}
S_0(r,\theta)=&-U\frac{\left(r^2+1\right)}{r} \cos \theta +U^3 \left[\left(\frac{6 r^2-1}{6
   r^5}-\frac{13}{6 r}\right) \cos \theta +\left(\frac{1}{2 r}-\frac{1}{6 r^3}\right) \cos 3 \theta \right]
   \\ &+U^5 \Bigg[\left(-\frac{7}{30 r^9}+\frac{3}{2 r^7}-\frac{43}{12 r^5}+\frac{35}{6 r^3}-\frac{479}{60 r}\right) \cos (\theta )+\left(-\frac{1}{36
   r^9}+\frac{4}{15 r^7}-\frac{3}{2 r^5}+\frac{43}{30 r^3}+\frac{19}{12 r}\right) \cos 3 \theta \\&\qquad \qquad \qquad\qquad\qquad +\left(\frac{7}{20 r^5}-\frac{1}{2 r^3}-\frac{1}{4 r}\right) \cos 5 \theta
   \Bigg] + \cdot\cdot\cdot.
 \end{align*}  
\end{widetext}

The boundary layer function becomes $\widehat S_0=S_0(1,\theta)$ and the maximum flow velocity is when $\cos \theta = 1$ ($\theta=\pi/2$). The corresponding maximum velocity is
\begin{widetext}
\begin{equation}
u_{0\rm max}=\frac{1}{r}\frac{\partial S_0(r,\theta)}{\partial \theta}\bigg|_{r=1,\theta=\frac{\pi}{2}}=2 U+\frac{7 }{3}U^3+\frac{176}{15}U^5+\frac{1511639}{18900}U^7+\frac{5084105183 }{7938000}U^9+\frac{311688814107079 }{55010340000}U^{11} + \cdot\cdot\cdot
\label{umax}
\end{equation}
\end{widetext}
This coincides with the expression for the velocity (in terms of Mach number) obtained  in \cite{rica01} via a Janzen-Rayleigh expansion applied to the classical problem of the flow of a compressible fluid passing around a solid disk.

The equation  \eqref{criterion} becomes hyperbolic beyond a critical velocity. It first  happens at $u_{\rm max}$ such that
\begin{equation}
1+U^2 = 3 u_{\rm max}^2.
\end{equation}
Solving this equation for $u_{\rm max}$  as shown in Eq.\eqref{umax} gives $U=0.263$, or in dimensionless units $v=0.37 c$. By considering the terms of the mainstream expansion up to $O(U^{40})$ we recover $v=0.36969(7)c$ in agreement with \cite{rica01}.

To get the analytical expression for the critical velocity for a finite size of the disk we need to consider 
the $\mathcal O (\varepsilon)$ contribution to the mainstream solution that satisfies the equations
\begin{eqnarray}
R_0 R_1 &=& - \nabla S_0 \cdot \nabla S_1\\
\left[1+U^2-3(\nabla S_0)^2\right]\nabla^2S_1&=&6 \nabla S_0\cdot\nabla S_1 \nabla^2 S_0.
\label{aamainS1}
\end{eqnarray}
For $S_1$ we employ the expansion similar to Eq.\eqref{S2d}, solve the ordinary differential equations for $S_{ij}$ to get
\begin{eqnarray}
S_1(r,\theta)=\frac{d_1}{r} U \cos \theta &+&U^3 \Bigg[\cos \theta  \left(\frac{{d_1}}{2 r^5}-\frac{2 {d_1}}{r^3}+\frac{{d_2}}{r}\right)\nonumber \\&+&\cos 3 \theta 
   \left(\frac{d_3}{r^3}-\frac{d_1}{2 r}\right)\Bigg]\cdot\cdot\cdot,
   \label{s1}
   \end{eqnarray}
where the constants of integration $d_i$ are found by matching the boundary-layer solution to \eqref{s1}. We substitute $r=1+\varepsilon\chi$ in \eqref{s1}, expand the solution in powers of $\varepsilon$ and match to the dominating linear in $\chi$  term in \eqref{S2}. The corresponding term in $\widehat{S_2}$ is expanded in powers of $v$ and in trigonometric functions. The resulting expression for the mainstream becomes
\begin{widetext}

\begin{eqnarray}
\frac{S_1(r,\theta)}{\sqrt{2}}&=&-U\frac{2 
  }{r} \cos \theta +U^3\Bigg[\left(\frac{5}{3 r^3}+\frac{1}{r}\right) \cos 3 \theta +\left(-\frac{1}{r^5}+\frac{4 }{r^3}-\frac{31}{3
   r}\right) \cos \theta \Bigg] 
   +U^5 \Bigg[\left(-\frac{17}{20 r^5}-\frac{3}{r^3}-\frac{5}{2 r^3}-\frac{1}{2r}\right) \cos 5 \theta\nonumber\\
   &+&\left(-\frac{5}{18 r^9}+\frac{53}{15 r^7}-\frac{16}{r^5}+\frac{1753}{60 r^3}+\frac{1}{r}+\frac{31}{6 r}\right) \cos 3 \theta  \nonumber \\ 
   &+&\left(-\frac{7}{3
   r^9}+\frac{37}{3 r^7}+\frac{5}{6 r^7}-\frac{65}{3 r^5}-\frac{31}{6 r^5}+\frac{106}{3
   r^3}-\frac{1161}{20 r}\right) \cos \theta  
 \Bigg] +\cdot\cdot\cdot
     \end{eqnarray}
   
   \clearpage
   
\end{widetext}
 The $\varepsilon $ term in the expansion for the maximum value of the velocity on the disk is
 \begin{eqnarray}
 u_{1\max}&=&\frac{1}{r}\frac{\partial S_1(r,\theta)}{\partial \theta}\bigg|_{r=1,\theta=\frac{\pi}{2}}\\
 &=&\sqrt{2}\Bigg[2U+\frac{46 U^3}{3}+\frac{8453 U^5}{60}+ \frac{5525323 U^7}{3780}\Bigg]+\cdot\cdot\cdot \nonumber
 \end{eqnarray}
 It is clear from this expression that  the maximum velocity on the surface of the obstacle is growing as the radius of the obstacle decreases for constant velocity of the mainstream. Therefore, the nucleation of the excitations on the surface of the object is not directly relevant to the maximum velocity for the objects of a finite radius as numerical simulations show (see Section 2). As we show in the next section the vortices and other excitations appear when it becomes energetically possible to create a doublet on the surface. The asymptotics for $\psi$ developed in this section allows us to get estimates of the energy of the system.

\subsection{ Critical velocity of nucleation}

The effect of the finite size of an obstacle on the critical velocity of vortex nucleation has been studied numerically \cite{huepe00}. It was demonstrated that the nucleation takes place when it becomes energetically favourable to create a vortex on the surface of the obstacle. Here we use this criterion to obtain the critical velocity of nucleation by analytical means.

It is convenient to consider a different rescaling of \eqref{modelone} in 
 units of $\vec x \to \xi \vec x$, $t \to (\xi^2 M /\hbar) t$, $v \to (\hbar /\xi M) U$ and $\psi \to (\psi_\infty e^{(-i U x)}) \psi$ with $\psi_\infty = (E/g)^{1/2}$ such that $\psi \to 1$ as $|\vec x| \to \infty$. The Eq.~\eqref{modelone} becomes
\begin{equation}\label{model}
2 i \partial_t \psi = - \Delta \psi + V \psi + (|\psi|^2-1) \psi + i 2 U \partial_x \psi.
\end{equation} 
 $V$ denotes the rescaled potential modeling the fixed impurities inserted into the fluid's flow. 
The energy of the system \eqref{model} is \cite{roberts}
\begin{equation}
E=\int |\nabla \psi|^2 + (1 -V- |\psi|^2)^2  d{\bf x}.
\label{energy}
 \end{equation}
 
 The solitary wave solutions such as vortex pairs and rarefaction pulses were analysed numerically in \cite{roberts} and asymptotically in \cite{berloff6}.  The lowest energy of vortical solutions is for the limiting case between a vortex pair and a rarefaction pulse: a doublet -- a single nodal point of $\psi$ when two vortices of opposite circulation collide. The doublet is moving through the uniform superfluid with velocity $U\approx0.45$. It's explicit form can be approximated by adapting the Pade approximations considered in \cite{berloff6}:
 \begin{eqnarray}
u_d&=&1 + \frac{a_{10}x^2 + a_{01} y^2-1}{1+c_{10}x^2 + c_{01}y^2 + c_{20} x^4 + c_{11} x^2 y^2 + c_{02} y^4}, \nonumber \\
v_d&=&\frac{x(b_{00} + b_{10}x^2 + b_{01} y^2)}{1+c_{10}x^2 + c_{01}y^2 + c_{20} x^4 + c_{11} x^2 y^2 + c_{02} y^4},
 \label{reim}
 \end{eqnarray}
 where $u_d(x,y)={\rm Re}(\psi)$ and $v_d(x,y)={\rm Im}(\psi)$ for the doublet.
 The far field expansions for $|{\bf x}|\rightarrow \infty$ were considered in \cite{roberts}:
 \begin{eqnarray}
u_d&\approx& 1 + m\frac{(2U-m)x^2-2 U(1 - 2 U^2)y^2}{2(x^2 + (1 - 2 U^2)y^2)^2},\nonumber \\
v_d&\approx& -\frac{m x}{x^2 + (1 - 2 U^2)y^2}.
 \end{eqnarray}
 To match it with \eqref{reim} we set
 \begin{eqnarray}
 a_{10}&=&\frac{1}{2}c_{20}m(2 U-m), \quad a_{01}=-c_{20} m (1-2 U^2) U, \nonumber \\b_{10}&=&-m c_{20},
 b_{01}=-mc_{20}(1 - 2 U^2), \\
 c_{11}&=& 2 c_{20} (1 - 2 U^2), \quad c_{02}=c_{20}(1 - 2 U^2)^2 \nonumber 
 \end{eqnarray}
  and determine $b_{00}, c_{20}, c_{10},c_{01}$ by expanding the stationary Eq. \eqref{model} with \eqref{reim} around zero and setting the constant term, the terms at $x^2$ and $y^2$ in  the real part of \eqref{model} as well as the term at $x$ in the imaginary part of \eqref{model} to zero. The known values of $U=0.45$ and $m=3.32$ complete the determination of unknowns in \eqref{reim}.
  
\begin{figure}
\epsfig{file=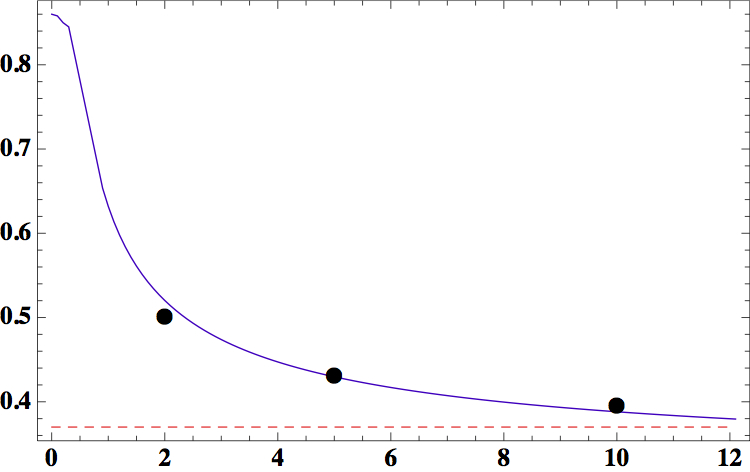,width=\columnwidth}\put(-262,143){ \textcolor{black}{$v_{\text{cr}}/c$}}\put(-108,2){ \textcolor{black}{$b/\xi$}}
\caption{Critical velocity of vortex nucleation  as a function of obstacle radius $b/\xi$. Solid line -- numerically determined critical velocity. The critical velocity $v_{cr}=0.36970c$ for an infinite obstacle is given by the dashed line. Dots -- analytically determined critical velocity as explained in the text.
 \label{velo}}
\end{figure}

  We approximate the wave function of the doublet sitting on the surface of the disk of the radius $b$  by 
  \begin{equation}
  \psi_d=u_d(x,y-b) + i v_d(x,y-b)
  \end{equation}
  and obtain the energy from \eqref{energy} where the integration is for $r>b$.
  The critical velocity of the vortex nucleation from the surface of the moving disk is then associated with the disk velocity at which the asymptotic solution \eqref{mt} reaches the energy of the doublet sitting on the surface of a stationary disk. Figure \ref{velo} summarizes our findings and compares the resulting critical velocities with the numerical solutions. We determined that the procedure for determining the criticality gives a good approximation for $20>b/\xi>2$, for large obstacles the criterion of  the velocity exceeding the local speed of sound becomes more accurate, whereas for the obstacle sizes of the order of the healing length the asymptotic expansion of the solution breaks down.
 
  In the next section we show that the shape of the potential modelling the impurity has a profound effect on the type of excitation  created.  

 

%
%
%
%

\section{2.  Nucleation of excitations: vortices and rarefaction pulses}

The nonlinear Schr\"odinger equation \eqref{modelone} possesses elementary excitations in the form of solitary waves: vortex pairs and rarefaction pulses -- finite amplitude sound waves \cite{roberts}. In 2d rarefaction pulses have lower energy and momentum than vortex pairs, so one may expect that small impurities, with radii smaller than healing length, will  generate rarefaction pulses rather than vortex rings \cite{pita}. It is also clear from the topology of the system that if the obstacle does not bring about the zero of the wave function of the condensate through the repulsive interactions the formation of a vortex pair always starts from a finite amplitude sound wave.  Therefore, one can envision that depending on the properties of the repulsive interaction induced by the impurity on the condensate different excitations are generated.

We start by modelling  the repulsive interactions between the obstacle at position $(x_i,y_i)$ and the condensate by  a potential 
\beq\label{pot}
V=V_0 \left(1- \tanh \left[(x - x_i)^2 + (y-y_i)^2 - b^2 \right] \right).
\eeq
 Here $V_0 > 0$ is the repulsive interaction strength between the impurity  and the condensate  and $b$ the impurity radius.


 \begin{figure}
\epsfig{file=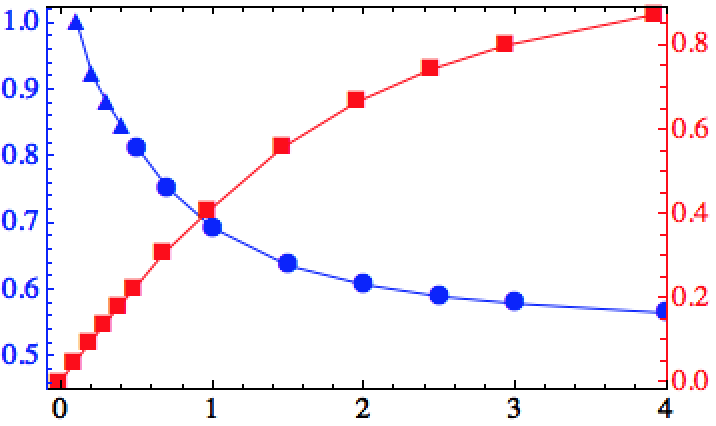,width=\columnwidth}\put(-256,128){ \textcolor{black}{$v_{\text{cr}}/c$}}\put(-93,2){ \textcolor{black}{$V_0$}}
\caption{ Critical velocity  at which  excitations are generated as function of obstacle height $V_0$ for fixed $b =\xi$. Dots represent the generation of vortex pairs, while triangles represent rarefaction pulses. Squares represent the depletion of the condensate  (uniform density minus the minimum of the density at the centre of the potential) due to the repulsive interactions with the obstacle \cite{mynote}.
 \label{hi}}
\end{figure}

Fig.\ref{hi} shows the dependence of the critical velocity on the height $V_0$ of the impurity potential. In particular we have numerically found that for a weaker potential (i.e., $V_0$ small) rarefaction pulses are generated rather than vortex pairs. In order  to distinguishing the generation of vortex pairs from the generation of the  rarefaction pulses we have throughout this work analyzed the excitations by evaluating if both the real and the imaginary part of the wave function are zero, when passing a radius of two healing lengths measured from the center of the impurity; setting a fixed radius is an unambiguous way to make an identification as for example a rarefaction pulse might gain energy when leaving the obstacle due to sound waves present in the condensate \cite{berloff6}, i.e., at different radii different excitations could be noticed in some circumstances.  We have observed that the smaller is the strength $V_0$ the greater is the critical velocity;  for zero depletion of the condensate the criticality agrees with the asymptotics considered in the previous section.

\begin{figure}[ht]
\begin{tabular}{ccc}
\begin{picture}(70,70)
\put(0,0) {\includegraphics[scale=0.2]{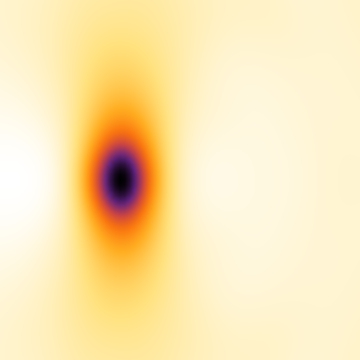} } \put(10,10){ \textcolor{black}{(a)}}
\end{picture}&
\begin{picture}(70,70)
\put(0,0) {\includegraphics[scale=0.2]{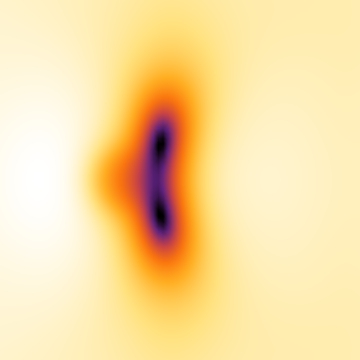} } \put(10,10){ \textcolor{black}{(b)}}
\end{picture}&
\begin{picture}(70,70)
\put(0,0) {\includegraphics[scale=0.2]{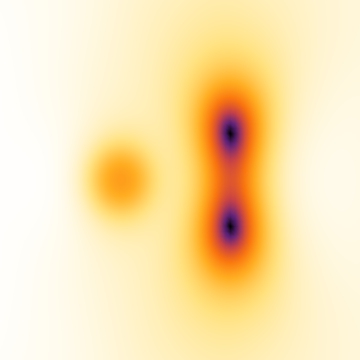} } \put(10,10){ \textcolor{black}{(c)}}
\end{picture}\\
\end{tabular}
\caption{Pseudo-color density plots of superfluid flow around an obstacle specified by $V_0 =0.7$, $b=1$ (measured in healing lengths) at $t=60$; (a), $t=84$; (b) and $t=90$; (c) with velocity $v=0.753 c$. The more luminous the picture the higher the density. Each picture shown corresponds to an area of $15\otimes15$   \cite{mymy}. The dimensionless units are used as specified in the main text.}
\label{fut}
\end{figure}

\begin{figure}[ht]
\begin{tabular}{ccc}
\begin{picture}(70,70)
\put(0,0) {\includegraphics[scale=0.2]{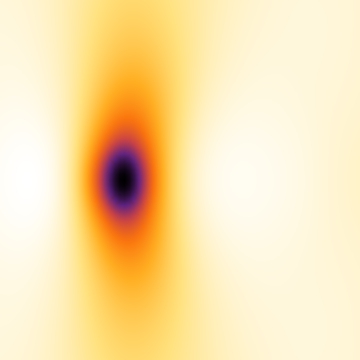} } \put(10,10){ \textcolor{black}{(a)}}
\end{picture}&\begin{picture}(70,70)
\put(0,0) {\includegraphics[scale=0.2]{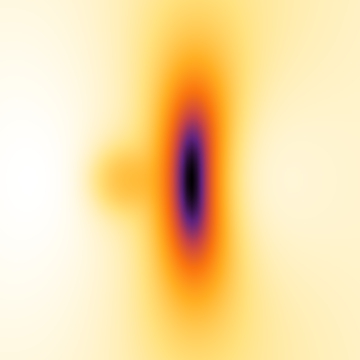} } \put(10,10){ \textcolor{black}{(b)}}
\end{picture}
&\begin{picture}(70,70)
\put(0,0) {\includegraphics[scale=0.2]{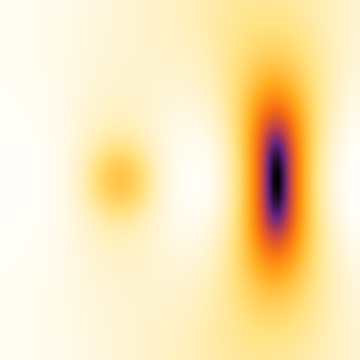} } \put(10,10){ \textcolor{black}{(c)}}
\end{picture}
\end{tabular}
\caption{Pseudo-color density plots of superfluid flow around an obstacle specified by  $V_0 =0.3$, $b=1$  at $t=150$; (a), $t=175.5$; (b) and $t=186$; (c) with flow velocity $v=0.88 c$. The more luminous the picture the higher the density. Each picture shown corresponds to an area of $15\otimes15$  \cite{mymy}. The dimensionless units are used as specified in the main text.}
\label{fut2}
\end{figure}

As mentioned above if the obstacle does not completely deplete the condensate density at its centre, then it is topologically impossible to create a vortex pair on non-zero background. Instead a finite amplitude sound wave is created at the condensate density minimum which can evolve into either a vortex pair or a rarefaction pulse as the wave separates from the obstacle and gains energy entering the bulk. The outcome in this case depends on the energetics created by the obstacle: the larger $V_0$ the more energetic solution emerges. This is illustrated in Figs.\ref{fut} and \ref{fut2}.
In Fig.\ref{fut} we show time snapshots illustrating  the emergence of a vortex pair for a stronger barrier.  A finite amplitude sound wave formed at the impurity atom $(a)$ evolves into a pair of vortices $(b)$ leaving towards the direction of the stream of the superfluid  $(c)$. Fig.\ref{fut2} illustrates the formation of a rarefaction pulse at a weaker obstacle; here after a finite amplitude sound wave is formed at the obstacle  it evolves into a rarefaction pulse that is carried away by the stream.

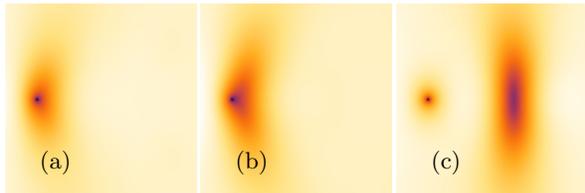
\begin{figure}[ht]
\begin{tabular}{ccc}
\begin{picture}(70,70)
\put(0,0) {\includegraphics[scale=0.2]{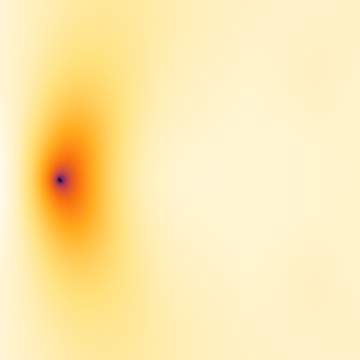} } \put(10,10){ \textcolor{black}{(a)}}
\end{picture}&
\begin{picture}(70,70)
\put(0,0) {\includegraphics[scale=0.2]{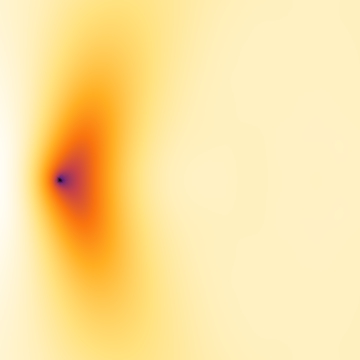} } \put(10,10){ \textcolor{black}{(b)}}
\end{picture}&
\begin{picture}(70,70)
\put(0,0) {\includegraphics[scale=0.2]{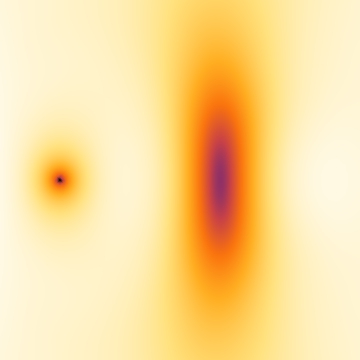} } \put(10,10){ \textcolor{black}{(c)}}
\end{picture}\\
\end{tabular}
\caption{Pseudo-color density plots of superfluid flow around a delta impurity at $t=21.5$ (a), $t=26.5$ (b), $t=41.5$ (c), with velocity $v=0.92c$. Each picture shown corresponds to an area of $12 \otimes 12$  \cite{my}. The more luminous the picture the higher the density. The dimensionless units are used as specified in the main text.}
\label{dir}
\end{figure}

\begin{figure}[ht]
\begin{tabular}{cc}
\begin{picture}(110,110)
\put(0,0) {\includegraphics[scale=0.25]{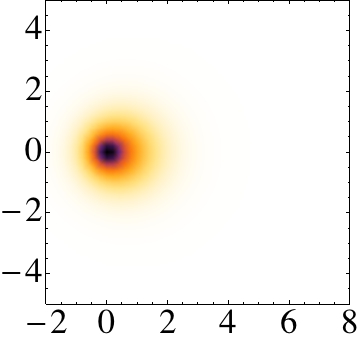} } \put(30,20){ \textcolor{black}{(a)}}
\end{picture}&
\begin{picture}(110,110)
\put(0,0) {\includegraphics[scale=0.25]{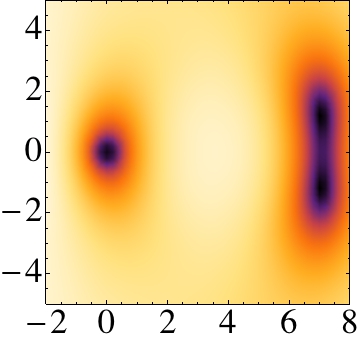} } \put(30,20){ \textcolor{black}{(b)}}
\end{picture}
\\\end{tabular}
\caption{Pseudo-color density plots of superfluid flow around a very small obstacle specified by $V_0=40$ and $\alpha=7$ at $t=0$ (a), $t=22.8$ (b) with velocity at infinity $v=0.92c$. Each picture shown corresponds to an area of $10 \otimes 10$  \cite{blue}. The more luminous the picture the higher the density. The dimensionless units are used as specified in the main text.}
\label{small}.
\end{figure}

{\it Delta-function impurity.} The special case of a single delta-function  impurity $V=\delta \left( x - x_i + y-y_i\right)$ can be regarded as a limiting case of the above setup, i.e., a single point obstacle at which the wave function is zero.  In this case at the critical velocity  slightly below the speed of sound rarefaction pulses are generated, see Fig.\ref{dir}, consistent with considerations in \cite{pita}.  However, for finite size obstacles  even smaller than the healing length  the  vortex pairs can be generated as soon as the the depletion of the condensate density is steep enough (even if the wave function is not zero at the obstacle), see Fig.\ref{small} where  we considered a potential of the form $
V=V_0 \left(1- \tanh \left[ \alpha ((x - x_i)^2 + (y-y_i)^2) \right] \right)$,
with $\alpha > 1$.

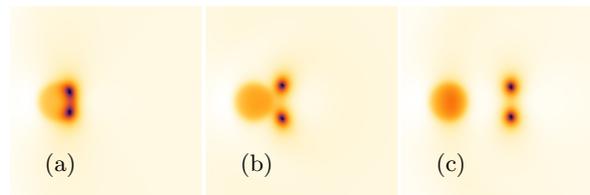
\begin{figure}[ht]
\begin{tabular}{ccc}
\begin{picture}(70,70)
\put(0,0) {\includegraphics[scale=0.2]{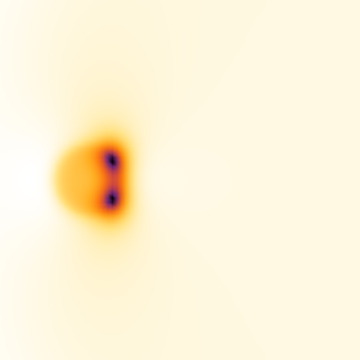} } \put(10,10){ \textcolor{black}{(a)}}
\end{picture}&
\begin{picture}(70,70)
\put(0,0) {\includegraphics[scale=0.2]{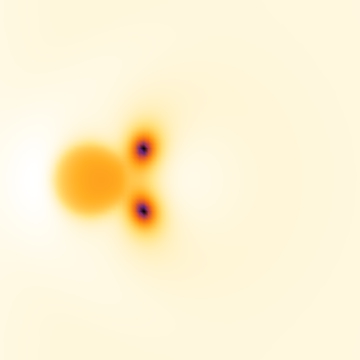} } \put(10,10){ \textcolor{black}{(b)}}
\end{picture}&
\begin{picture}(70,70)
\put(0,0) {\includegraphics[scale=0.2]{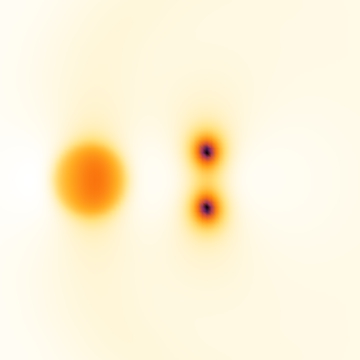} } \put(10,10){ \textcolor{black}{(c)}}
\end{picture}\\\end{tabular}
\caption{Pseudo-color density plots of superfluid flow around a obstacle specified by $V_0=0.25$ and $b=4$ at $t=57$; (a), $t=75$; (b), $t=105$; (c) with velocity at infinity $v=0.8c$. Each picture shown corresponds to an area of $40 \otimes 40$ \cite{my}. The more luminous the picture the higher the density. The dimensionless units are used as specified in the main text.}
\label{bigga}.
\end{figure}

Our numerical analysis shows that a stronger dip in the condensate makes a vortex pair favorable, while less deep depletions of small radius favor rarefaction pulses. In particular for weak BEC-impurity interactions of bigger radius $(V_0 \ll 1, b \gg 1)$ we have found that vortex pairs are generated, see  Fig.\ref{bigga}. Hence, it depends on the energy put into the system via the potential which excitation can be afforded, i.e., low energy potentials favour energetically cheaper rarefaction pulses while higher energy potentials more expensive vortex pairs.

\begin{figure}[ht]
\begin{tabular}{ccc}
\begin{picture}(70,70)
\put(0,0) {\includegraphics[scale=0.2]{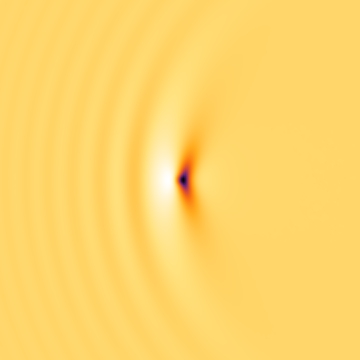} } \put(10,10){ \textcolor{black}{(a)}}
\end{picture}&
\begin{picture}(70,70)
\put(0,0) {\includegraphics[scale=0.2]{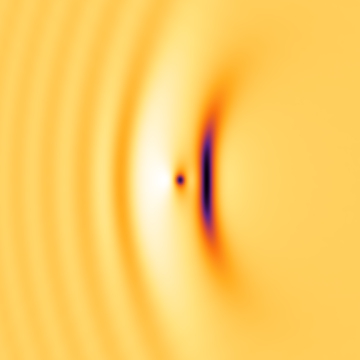} } \put(10,10){ \textcolor{black}{(b)}}
\end{picture}
&\begin{picture}(70,70)
\put(0,0) {\includegraphics[scale=0.2]{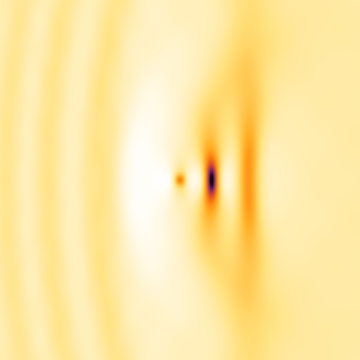} } \put(10,10){ \textcolor{black}{(c)}}
\end{picture}
\end{tabular}
\caption{Pseudo-color density plots of superfluid flow around a obstacle specified by $V_0=0.2$ and $b=1$ at $t=45$; (a), $t=82.5$; (b) and $t=180$; (c) with Mach number $v=1.01c$. The more luminous the picture the higher the density. Each picture shown corresponds to an area of $35 \otimes 35$  \cite{supii}. The dimensionless units are used as specified in the main text.}
\label{supersonic}
\end{figure}

{\it Supersonic flow,-} Finally, for completion,  we consider the superfluid at supersonic speed (i.e., for Mach numbers 
$M= v/c > 1$) passing a small and only weak obstacle, that does not lead to a complete depletion of the condensate density at its position. The case of a strongly interacting (delta function) obstacle has been considered in \cite{glad,patt}, where oblique dark soliton trains were observed in the wake of the obstacle.  Here in Fig.\ref{supersonic} we present the emergence of rarefaction pulses in the stream of the condensate accompanied by Cherenkov waves outside the Mach cone. The rarefaction pulse in the wake of the obstacle clearly differs from the solitary waves (or continuous stream of vortex pairs) spotted for heavy (delta function) impurities in \cite{glad} in so far as not a full depletion of the condensate has been observed.

\section{3. Superfluid regimes in the lattice of impurities}

In this section we consider $N$ inserted obstacles at positions $(x_i,y_i)$ with $i \in \{1,\ldots, N \}$ that generate an external  potential of the form
\beq\label{latte}
{\cal V}=V_0 \sum^N_{i=1}\left(1- \tanh \left[(x - x_i)^2 + (y-y_i)^2 - b^2 \right] \right).
\eeq
The potential function of the whole array ${\cal V}$ depends on the spatial order of the inserted impurities, the radius $b$ of the atoms and the strength of BEC-atom interaction $V_0$. We suppose that there are no interactions between the impurities themselves and that they are stationary, but remark that the additional presence of an atom in the superfluid leads to a change of the potential generated by the other atoms -- their states are squeezed \cite{tim} within the superfluid. In this work, however, we won't address the issue {\it how} the size of the impurities materializes, but elucidate that {\it having a certain size} effects the superfluid in a certain way.

\subsection{Impurities arranged into regular lattices}

\begin{figure} 
\epsfig{file=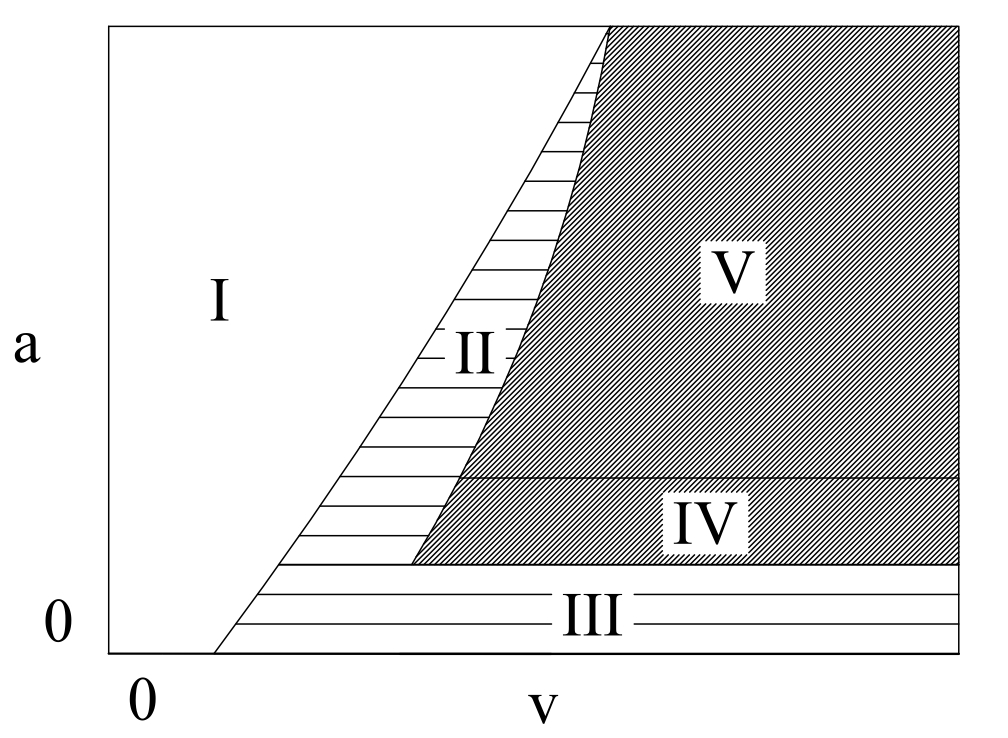,width=\columnwidth}
\caption{The distinct regions in superfluid flow around regular lattices on a rectangular area. Here $a$ denotes the distance between nearest neighbors and $v$ the absolute value of the superfluid velocity  at infinity. Hatched areas ${\rm II}$, ${\rm III}$ show regions where excitations are outside the lattice and ${\rm V}$, ${\rm IV}$  indicate the regimes where excitations are within the lattice as well.
\label{graf}}
\end{figure}

Let us consider the superfluid's flow around impurities arranged on a triangular lattice within a finite sized  rectangular area. We  distinguish different regimes in superfluid flow passing the lattice that besides the characteristics of impurity atoms $V_0$ and $b$ depend on the velocity $v$, the distance between nearest neighbors $a$ and the size of the lattice. The various regimes are presented in Fig.\ref{graf} qualitatively as functions of $v$ and $a$: Area ${\rm I}$ corresponds to the superfluid phase, i.e., no excitations emerge in the superfluid's stream. The regime ${\rm II}$ is characterized by the emergence of vortices from the boundary of the network. ${\rm III}$  describes very dense packing of impurity atoms, such that for those velocities vortices are created on the boundaries of the entire lattice, { while the superfluid is expelled from within the lattice}. Region ${\rm IV}$ is described by excitations within the lattice, { which span from excitations smaller then the healing length and ones bigger} than a few healing lengths, i.e., emergent macroscopic structures. In region ${\rm V}$ vortices or rarefaction pulses are generated within the lattice as well as outside the lattice without forming bigger structures as a consequence of the sparsity of the impurity atom distribution.

\begin{figure}
\hbox{\vspace{3mm}}
\hbox{
\begin{picture} (100,60)
 \put(0,0){\includegraphics[width=248pt, height=62pt]{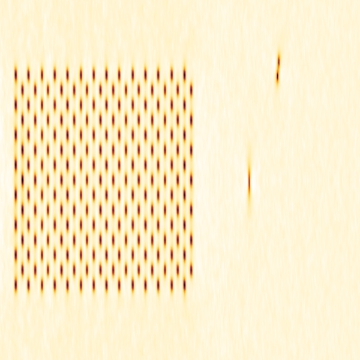} }  \put(220,45){ \textcolor{black}{(a)}} \put(10,30){ }
\end{picture} }
\hbox{
\begin{picture} (100,60)
 \put(0,0){\includegraphics[width=248pt, height=62pt]{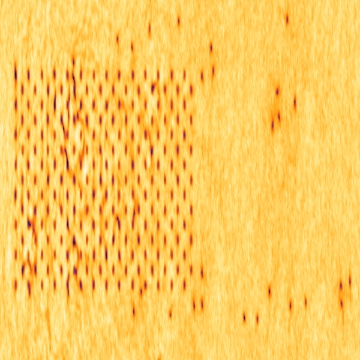} } \put(220,45){ \textcolor{black}{(b)}} \put(10,30){ }
\end{picture}
 }
\caption{Pseudocolor density plots $|\psi|^2(x,y)$ of qualitatively distinguishable phases in superfluid flow passing a hexagonal lattice of impurities. Shown is the condensate density at velocity $v=0.5c$ (a) and   $v=0.555c$ (b).  Other parameters of the system: Array size $A=240\otimes72$, frame shown $480\otimes 120$, $a=5$, $b=1.5$,$V_0 =1$, all snapshots at  $t=125$ \label{phases}\cite{mynote2}. The more luminous the picture the higher the density. The dimensionless units are used as specified in the main text.}
\end{figure}
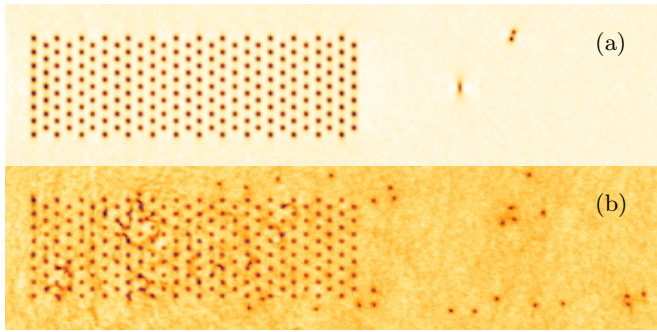

\begin{figure}
\epsfig{file=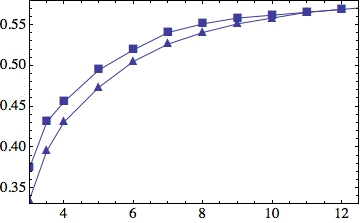,width=\columnwidth}\put(-260,122){ \textcolor{black}{$v_{\text{cr}}/c$}}\put(-90,2){ \textcolor{black}{$a$}}
\caption{Critical velocities as functions of the distance between impurity to its nearest neighbors $a$. For this comparison the remaining free parameters have been chosen to be $V_0=1$, $b=1.5$ while the area at which the triangular lattice of atoms has been present is $200\otimes120$ (measured in healing lengths). Triangles represent data points of the first and squares  data points of the second critical velocity. The interpolation between data points is linear. \cite{mynote2}.
 \label{velo2}}
\end{figure}

In Fig. \ref{phases} we present the superfluid flow towards the right hand side around a sparse triangular lattice in the regions ${\rm II}$ and ${\rm V}$.  Above the first critical velocity  $v_1$, excitations are generated at the end of the array and directly move into the wake (b) (region ${\rm II}$). Above the second critical velocity $v_2$ excitations are continuously generated within the lattice  and move - carried by  the fluids stream - through the impurity network towards the wake of the system (c) (region ${\rm V}$).  In Fig.\ref{velo2} we show  the first  $v_1$ and second $v_2$ critical velocities as functions  of the mean distance between nearest neighbors $a$ at the triangular lattice.

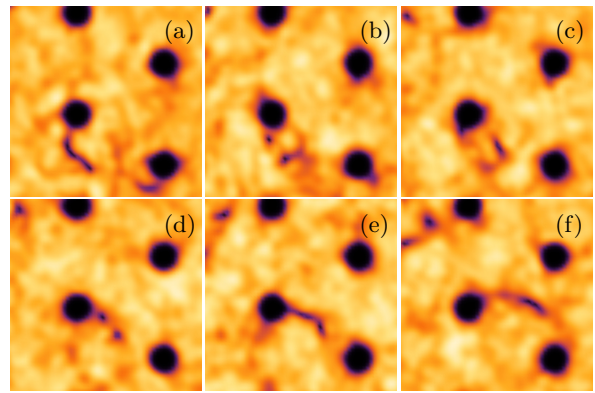
\begin{figure}[ht]
\begin{tabular}{ccc}
\begin{picture}(70,70)
\put(0,0) {\includegraphics[scale=0.2]{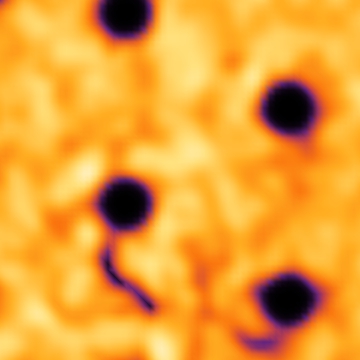} } \put(55,60){ \textcolor{black}{(a)}}
\end{picture}&
\begin{picture}(70,70)
\put(0,0) {\includegraphics[scale=0.2]{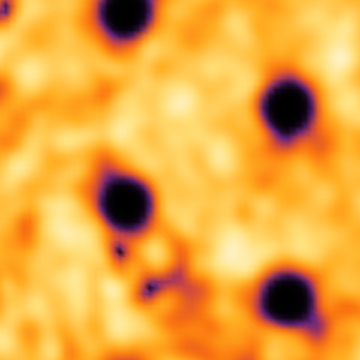} } \put(55,60){ \textcolor{black}{(b)}}
\end{picture}&
\begin{picture}(70,70)
\put(0,0) {\includegraphics[scale=0.2]{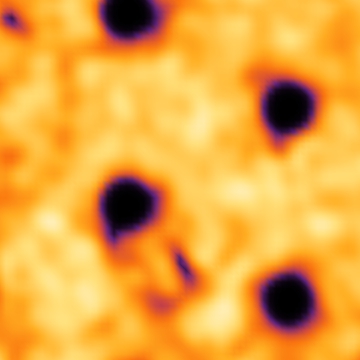} } \put(55,60){ \textcolor{black}{(c)}}
\end{picture}\\
\begin{picture}(70,70)
\put(0,0) {\includegraphics[scale=0.2]{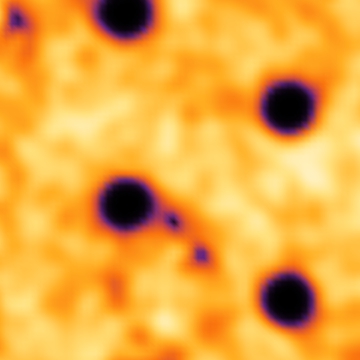} } \put(55,60){ \textcolor{black}{(d)}}
\end{picture}&
\begin{picture}(70,70)
\put(0,0) {\includegraphics[scale=0.2]{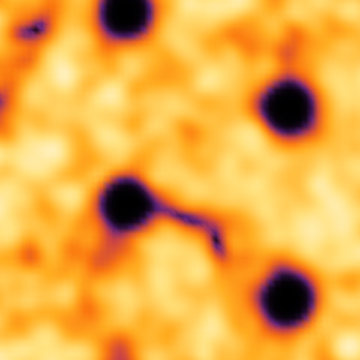} } \put(55,60){ \textcolor{black}{(e)}}
\end{picture}&
\begin{picture}(70,70)
\put(0,0) {\includegraphics[scale=0.2]{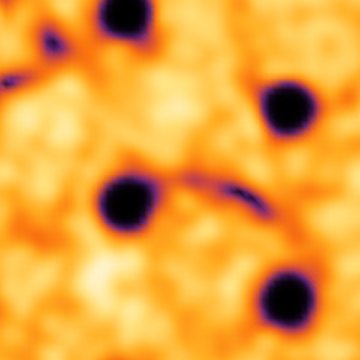} } \put(55,60){ \textcolor{black}{(f)}}
\end{picture}\\
\end{tabular}
\caption{Pseudo-color density plots of superfluid flow around a obstacles on arranged on a triangularlattice specified by $V_0=10$ and $b=2$ at $t=310$; (a), $t=312.5$; (b), $t=315$; (c); $t=317.5$; (d), $t=320$; (e), $t=325$; (f) with velocity at infinity $v=0.67c$. Each picture shown corresponds to an area of $30 \otimes 30$  \cite{mynote34}. The more luminous the picture the higher the density. The dimensionless units are used as specified in the main text.}
\label{pair}.
\end{figure}

Let us take a closer look on the lattice dynamics in region {\rm V} corresponding to a lattice of intermediate density of impurities. We could identify different processes of three body interactions of vortex dipoles with single vortex among that are the {\it flyby regime} and the {\it reconnection regime} \cite{Parker}. The flyby characterizes the scenario where an incoming  vortex pair gets deflected by the third vortex and the reconnection regime describes the situation where vortex $1$ of the pair is coupled to the third vortex and the other vortex $2$ of the pair is left behind. Moreover the transition from vortex pair to rarefaction pulse and vice versa due to loss or gain of energy through sound waves could be identified.
In Fig.\ref{pair} an example of a vortex pair acquires  a velocity component  transversal to the motion of the fluid by pinning to the potential spikes is shown (see pictures (a),(b),(c),(d),(e),(f)).
Here a vortex pair  looses one vortex to an impurity. As the vortex moves away from the obstacle it pulls the trapped vortex back,  while  acquiring an additional transversal velocity component and heading towards the next impurity. This process can be classified as a {\it flyby}, when applying a vortex mirror-vortex analogy \cite{Mason}.
In addition we could identify a locally stationary single vortex placed at the centroid of the triangle formed by nearest neighbor impurities. When other vortices enter the scene they almost form a triangle by connecting with the impurity atoms at the corners ({\it connecting regime}) and due to gain of energy through perturbations subsequently decay into many vortices. 
\begin{figure}[ht]
\hbox{
\begin{picture} (220,110)
 \put(10,-10){\includegraphics[width=220pt, height=110pt]{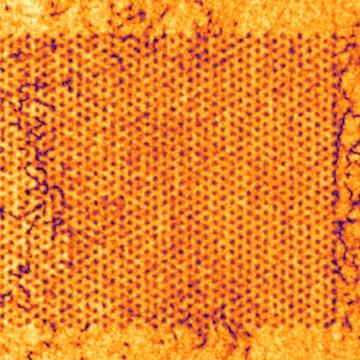} } \put(190,90){ \textcolor{white}{(a)}}
\end{picture} }
\begin{tabular}{cc}
\begin{picture}(110,110)
\put(0,-10) {\includegraphics[scale=0.3]{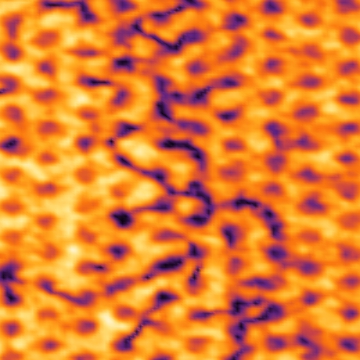} } \put(90,90){ \textcolor{white}{(b)}}
\end{picture}&
\begin{picture}(110,110)
\put(0,-10) {\includegraphics[scale=0.3]{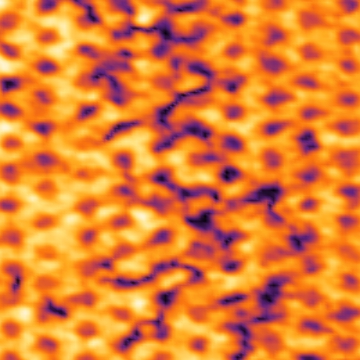} } \put(90,90){ \textcolor{white}{(c)}}
\end{picture}
\end{tabular}
\caption{Pseudo-color density plots $|\psi|^2(x,y)$ of superfluid flow passing dense hexagonal distributed atoms with velocity $v=0.5c$. The size of atom array is $290 \otimes 180$, the frame size shown in picture (a) is $220 \otimes 220$ at $t=522.5$. Picture (b) is a detail  of frame size $70 \otimes 70$ at $t=522.5$ and picture (c) at $t=525$ (with frame size $70 \otimes 70$). The atom parameters were chosen to be $V_0 =1$, $a=3$ and $b=1.5$. \cite{mynote34}. The more luminous the picture the higher the density. The dimensionless units are used as specified in the main text. \label{inside2}}
\end{figure}
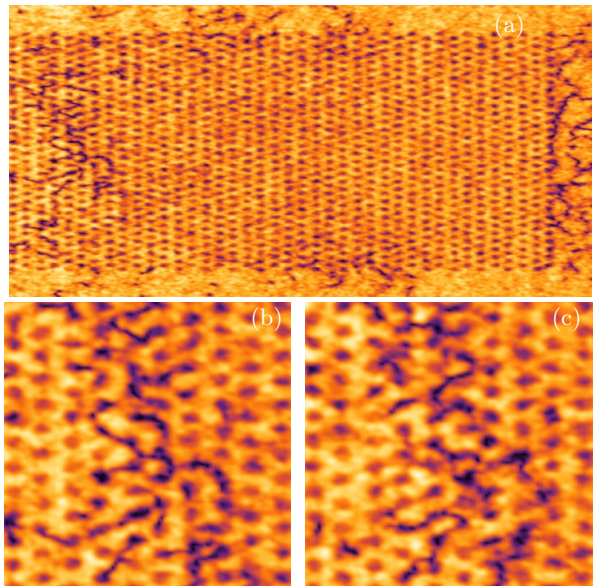
For weaker BEC-impurity interactions we have found an intermediate regime, where  (for example for parameters $V_0 =0.1$ and $b=1$) vortices are generated although single impurities of the same specification would prefer the generation of rarefaction pulses; high energy rarefaction pulses absorb energy due to perturbations present in the lattice and once their energy passes some threshold they transform into an energetically favorable vortex pair \cite{berloff6}. For {\it very weak} BEC-impurity interactions ($V_0$ is very small) solely rarefaction pulses are generated in the stream of the condensate within and at the wake of the impurity lattice, which are slightly deflected as they pass weak potential spikes.

The region {\rm IV} in Fig.\ref{graf} is characterized by high density arrays of impurities, which yield excitations that span over several neighboring impurities. In Fig.\ref{inside2} we show snakes of excitations moving through the lattice. These excitations are either zeros of wave function and therewith can be identified as vortices or are more comparable with rarefaction pulses not fully depleting the condensate. In particular we have observed that excitations within the lattice might move in opposite direction to the mainstream direction. With more narrow  space between impurities further excitations  are present within the lattice. As there is not enough space for a fully developed vortex pair or rarefaction pulse between neighboring impurity atoms, excitations in such lattice emerge as finite amplitude sound waves, i.e, spontaneously occurring density depletions between two neighboring atoms, which occasionally persist and move within the lattice generally towards all possible directions.

\begin{figure}
\epsfig{file=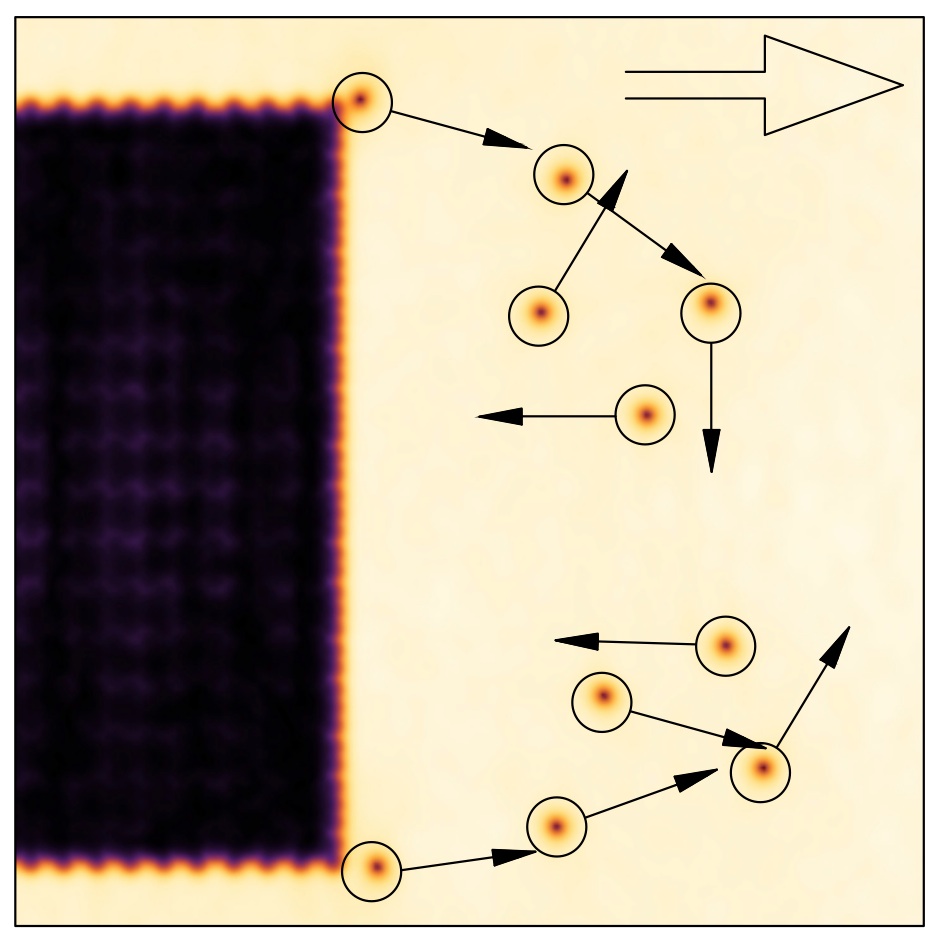,width=50mm}
\caption{Pseudo-color density plots $|\psi|^2(x,y)$ of superfluid flow passing very dense triangular lattice of impurities with velocity $v=0.2c$. The frame size shown is $150 \otimes 150$ at $t=539.25$. The impurity parameters were chosen to be $V_0 =1$, $a=1.6$ and $b=1.5$. \cite{mynote34}. The more luminous the picture the higher the density. Arrows point towards the direction of motion of the encircled vortices, and the big arrow indicates the direction of the superfluid mainstream. The dimensionless units are used as specified in the main text. \label{slip}}
\end{figure}

 Finally we have considered very high densities of impurity atoms with $V_0$ large enough such that the condensate is (almost) expelled from the lattice and the velocity is chosen such that the superfluid phase is surpassed, i.e., {\rm III} in Fig.\ref{graf}. Here we have found that vortices are generated at the boundaries of the lattice. In the wake some vortices enter the {\it slipstream region}, such that no significant motion between vortices and lattice is present or movement towards the lattice might occur - an analog situation as encountered for moving obstacles generating turbulent flow in normal fluids. In Fig.\ref{slip} we indicate the motion of the vortices in the slipstream region for a very slow superfluid with only few vortices present in the wake, i.e., two counter propagating curls of vortices evolving from both edges of the lattice.

\subsection{Uniformly distributed impurities}

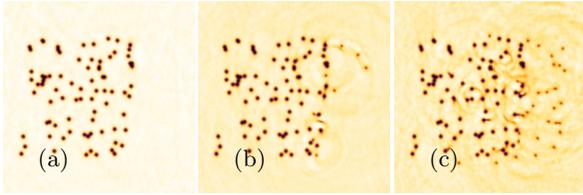
\begin{figure}[ht]
\begin{tabular}{ccc}
\begin{picture}(70,70)
\put(0,0) {\includegraphics[scale=0.2]{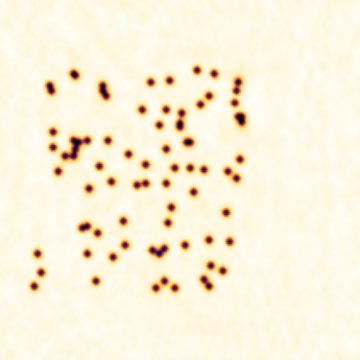} } \put(10,10){ \textcolor{black}{(a)}}
\end{picture}&
\begin{picture}(70,70)
\put(0,0) {\includegraphics[scale=0.2]{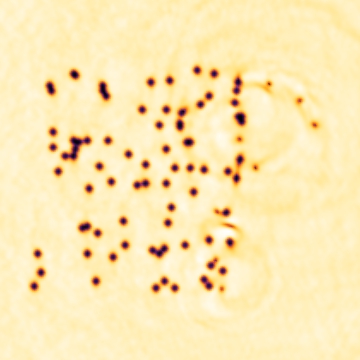} } \put(10,10){ \textcolor{black}{(b)}}
\end{picture}&
\begin{picture}(70,70)
\put(0,0) {\includegraphics[scale=0.2]{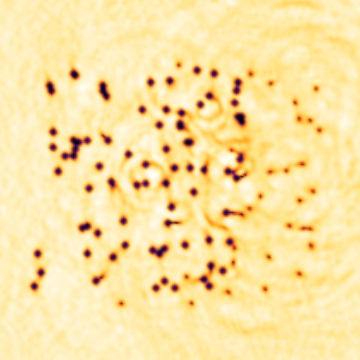} } \put(10,10){ \textcolor{black}{(c)}}
\end{picture}\\
\end{tabular}
\caption{Pseudo-color density plots $|\psi|^2(x,y)$ of phases in superfluid flow passing uniformly distributed atoms with different velocities at $t=165$. The atom number is $N_{\mathcal A} =80$,  size of possible atom positions $120 \otimes 130$ (measured in healing length), the frame size shown is $210 \otimes 210$ in same units and the atom parameters were chosen to be $V_0 =2$ and $b=2$. Picture (a) shows the flow at $v=0.2c$, (b) at $v=0.32c$, (c) at $v=0.39c$. All pictures show the same (random) distribution of atoms and represent a snapshot at $t=165$ \cite{mynote22}. The more luminous the picture the higher the density. The dimensionless units are used as specified in the main text. \label{ranpha}}
\end{figure}

We now turn to the regimes of a flow passing randomly distributed impurities occupying a finite area $A$. These inserted atoms can be regarded as an ideal gas of $N_{\mathcal A}$ uniformly distributed noninteracting particles in the plane $\mathbb R^2$ at an instant of time. We denote the density of particles (or impurities) given by the total particle number per area by $n$. To determine the mean distance between particles we recognize that the probability of finding another particle within the distance $r$ from its origin is $P_{[r,r+dr]}= 2 \pi r n dr$. The probability to find a particle outside the disc, i.e., in $[r,\infty]$, is $P_{[r,\infty]} = 1- \pi r^2/A$, where $A$ is the total area.
Hence the probability distribution function of the distance to the nearest neighbor is
\beq\label{one}
P_N(r) = 2 \pi r N/A \left( 1- \pi r^2/A \right)^{N-1},
\eeq
which for $n$ fixed becomes in the limit $N \to \infty$
\beq\label{infinite}
P(r) = 2 \pi r \exp(- \pi r^2n) n .
\eeq
Note that $\eqref{infinite}$ might be regarded as a good approximation to $\eqref{one}$ for large $N$.
Thus, the mean distance $\overline a$ is given  by considering the expectation value
\beq\label{exp}
\overline a \equiv \mathbb E_1 [r] = \int^\infty_0 r 2 \pi \exp(- \frac{\pi r^2}{n}) n dr =  \frac{1}{2 \sqrt{n}}.
\eeq
In this sense a random distribution of particles determined by its area and number relates to the mean distance, i.e.,  $n = N_{\mathcal A}/A= \frac{1}{4 \overline a^2}$. In Fig.\ref{ranpha} we present results showing that even for systems of less chosen structure than fixed lattices, qualitatively different phases can be distinguished. That is a phase of dissipationless superfluid flow, the generation of first excitations in the wake of the lattice and generation of excitations within the lattice.  Figure \ref{ranpha}  shows the density of the condensate for  the superfluid flow around inserted impurities uniformly distributed on a finite domain \cite{mynote2}, \cite{mynote3}. Fig.\ref{ranpha} $(a)$ shows a superfluid's flow without dissipation of energy and generation of elementary excitations, (b) corresponds to  a flow above criticality carrying excitations in the wake of the superfluid and (c) an even faster flow at which excitations are generated within the array.

\section{Conclusions}
The generation of vorticity by a moving superfluid has generated a lot of experimental and theoretical work.  In our paper we re-examine this problem by developing an asymptotic and analytical methods for finding the flow around an obstacle and for determining the critical velocity of vortex nucleation. We numerically study the various excitations generated above the criticality. We determine the regimes when a vortex pair or a finite amplitude sound wave is generated depending on the energetics of the obstacle.
We described several novel regimes as a superfluid flows  an array of impurities motivated by recent experiments. 

\section{Acknowledgements}
 F.P.  is financially supported by the UK Engineering and Physical Sciences Research Council (EPSRC) grant EP/H023348/1 for the University of Cambridge Centre for Doctoral Training, the Cambridge Centre for Analysis and a KAUST grant.


\begin{thebibliography}{10}
\bibitem{mynote} We numerically generated solutions  using fourth order finite differences scheme in space and the fourth order Runge-Kutta time integration on a computational window of $361$ space units in $x$ and $121$ units along the direction $y$ transverse to the superfluid flow. Typical space steps are about $\Delta x=0.25$ and time steps $\Delta t =  10^{-3}$. At the boundaries we have implemented an absorption layer of 20 grid points to minimise the reflection of emitted waves.
\bibitem{mymy} We have considered a grid of $350$ units in x direction and $200$ in the transverse direction. Step sizes in both directions have been $1/5$ healing layer and time steps have been $\Delta t = 0.00015$.
\bibitem{my} We have considered a Dirichlet-type obstacle at a computational grid of $700$ units in x direction and $400$ in the transverse direction. Step sizes in both directions have been $1/10$ healing layer  and time steps have been $\Delta t = 0.00005$.
\bibitem{blue} We have considered a computational grid of $500$ units in x direction and $500$ in transverse direction. Step sizes in both directions have been $1/5$ healing layer  and time steps have been $\Delta t = 0.00002$.
\bibitem{supii} The grid was $500$ units in x direction and $500$ in the transverse direction. Step sizes in both directions have been $1/4$ healing layer and time steps have been $\Delta t = 0.0001$.
\bibitem{mynote2} The computing grid has been chosen to be $440$ points in $x$ direction and $200$ points in $y$ direction. Time steps were about $\Delta t=0.004$ and space steps $\Delta x=1$. At the boundaries of the grid we have implemented an absorption layer of $14$ points in width to simulate open boundary conditions.
\bibitem{mynote34} The computing grid has been chosen to be $900$ points in $x$ direction and $500$ points in $y$ direction. Time steps were about $\Delta t=0.00025$ and space steps $\Delta x=0.5$. At the boundaries of the grid we have implemented an absorption layer of $14$ points in width to simulate open boundary conditions.
\bibitem{mynote3} Our simulations on randomly distributed atoms relied on the Fortran RANDOM NUMBER routine. In addition to implement different seeds for different lattices of randomly distributed atoms each time a new scenario was considered we have utilized the Fortran DATE AND TIME subroutine. Based on this routine our program generating a random seed depended on the milliseconds, second, and minutes of a particular date, hence ensuring randomness in so far as the likelihood of choosing the same seed has been vanishing in our setting.
\bibitem{mynote22} The computing grid has been chosen to be $400$ points in $x$ direction and $250$ points in transversal direction. Time steps were about $\Delta t=0.00015$ and space steps $\Delta x=1$. At the boundaries of the grid we have implemented an absorption layer of $14$ points in width.
\end{thebibliography}
\end{document}